# High-Mobility Carriers Induced by Chemical Doping in the Candidate Nodal-Line Semimetal CaAgP

Yoshihiko Okamoto,[1,*] Kazushige Saigusa,[1] Taichi Wada,[1] Youichi Yamakawa,[2] Ai Yamakage,[2] Takao Sasagawa,[3] Naoyuki Katayama,[1] Hiroshi Takatsu,[4] Hiroshi Kageyama,[4] and Koshi Takenaka[1]

[1]*Department of Applied Physics, Nagoya University, Nagoya 464-8603, Japan*
[2]*Department of Physics, Nagoya University, Nagoya 464-8602, Japan*
[3]*Laboratory for Materials and Structures, Tokyo Institute of Technology, Yokohama 226-8503, Japan*
[4]*Graduate School of Engineering, Kyoto University, Kyoto 615-8510, Japan*

We report the electronic properties of single crystals of candidate nodal-line semimetal CaAgP. The transport properties of CaAgP are understood within the framework of a hole-doped nodal-line semimetal. In contrast, Pd-doped CaAgP shows a drastic increase of magnetoresistance at low magnetic fields and a strong decrease of electrical resistivity at low temperatures probably due to weak antilocalization. Hall conductivity data indicated that the Pd-doped CaAgP has not only hole carriers induced by the Pd doping, but also high-mobility electron carriers in proximity of the Dirac point. Electrical resistivity of Pd-doped CaAgP also showed a superconducting transition with onset temperature of 1.7-1.8 K.

## I. INTRODUCTION

In recent years, nodal-line semimetals have been intensively studied as a system for realizing topologically nontrivial electronic states. A nodal line is formed by the intersection of linearly dispersive bands in momentum space. One of the first examples was the band contact line in graphite [1,2]. Subsequent studies showed that various topological phenomena unique to nodal lines, such as flat Landau levels and a topological surface band, are realized in materials containing nodal lines [2-4]. This surface band has a drumhead shape that can give rise to a large density of states, when a nodal line surrounds a time-reversal invariant momentum [5-8]. Materials with a nodal line at the Fermi energy ($E_F$) and in particular, nodal line semimetals, where the only nodal line is at $E_F$, are expected to exhibit not only topologically nontrivial electronic states, but also various interesting transport properties unique to nodal lines, such as three-dimensional quantum Hall effect and topological surface superconductivity [9-14].

Until now, various nodal-line materials have been experimentally studied. For example, angle-resolved photo-emission spectroscopy (ARPES) experiments on single crystals of ZrSiS and PbTaSe$_2$ have demonstrated the presence of nodal lines [15-18]. However, the nodal lines in these materials considerably deviated from $E_F$ and/or normal bands crossed $E_F$. Therefore, the properties of nodal lines are difficult to ascertain from their transport properties. For CaAgAs, experiments using single crystals showed the presence of a torus-shaped Fermi surface reflecting a ring-shaped nodal line, i.e., a nodal ring [19-23]. Since CaAgAs does not have other Fermi surfaces of normal bands, the influence of the nodal ring can appear significantly in its electronic properties. According to first-principles calculations, however, a band gap of ~0.1 eV opens at the nodal ring due to the strong spin-orbit coupling of As, suggesting that CaAgAs is more interesting as a topological insulator [6]. A high-temperature phase of Ca$_3$P$_2$ is an ideal candidate for a nodal-line semimetal, because it has a nodal ring at $E_F$ and the size of the spin-orbit gap is much smaller than that of CaAgAs [7,24]. Unfortunately, physical properties of Ca$_3$P$_2$ have not been uncovered, mainly due to chemical instability. Thus, not only electronic properties unique to the nodal line, but also basic transport properties of nodal-line materials are untapped experimentally.

Here, we report the transport properties of single crystals of undoped and Pd-doped CaAgP, which realized intriguing electronic properties caused by the nodal line. CaAgP is a 3$p$ analogue of CaAgAs and crystallizes in the hexagonal ordered-Fe$_2$P type structure [25]. First-principles calculations without spin-orbit coupling indicated that linearly dispersive bands cross at $E_F$, forming a nodal ring surrounding the Γ point [6]. Even when the spin-orbit coupling is switched on, the size of the spin-orbit gap is on the order of 1 meV, much smaller than that in CaAgAs, suggesting that CaAgP is more promising as a nodal-line semimetal. In previous experimental studies on CaAgP, transport properties of polycrystalline samples and ARPES on a single crystal were measured [26,27]. In the former study, $E_F$ of the polycrystalline sample was located far below (0.4 eV) the Dirac point probably due to lattice defects, and the samples showed conventional metallic resistivity and magneto-resistance. The latter study, which combined the ARPES measurements and calculation, suggested that band inversion does not occur in CaAgP.

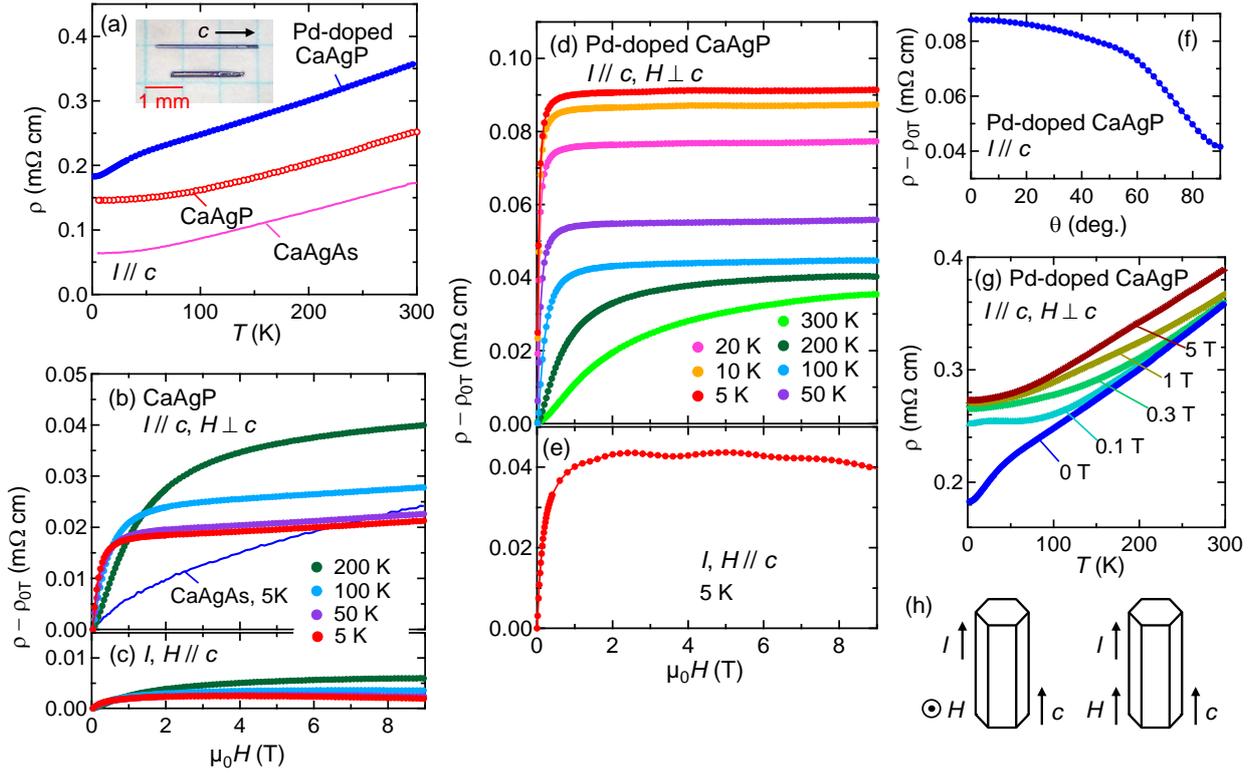

Fig. 1. Electrical resistivity of single crystals of undoped and Pd-doped CaAgP and CaAgAs. (a) Temperature dependence of electrical resistivity. The inset shows the single crystals of the undoped (upper) and Pd-doped (lower) CaAgP. (b-e) Magnetic field dependence of electrical resistivity for the undoped (b,c) and Pd-doped (d,e) CaAgP measured at various temperatures. Transverse and longitudinal magnetoresistances are shown in (b,d) and (c,e), respectively. (f) Magnetic-field orientation dependence of magnetoresistance for the Pd-doped CaAgP at 5 K. Angles of 0° and 90° correspond to the perpendicular and parallel directions relative to the electric current, respectively. In (b-f), the differences between electrical resistivity and zero-field resistivity are shown. (g) Temperature dependence of electrical resistivity for the Pd-doped CaAgP measured at various magnetic fields. The magnetic fields were applied perpendicular to the electric current. (h) Schematic pictures showing the directions of electric current and magnetic field. The left panel shows the cases of (b), (d), (g), and 0° in (f) and the right panel shows those of (c), (e), and 90° in (f).

## II. EXPERIMENTS

The single crystals of undoped and Pd-doped CaAgP were prepared by a flux method using Bi as a flux. Molar rations of 1.1:1:0:1 and 1.1:0.9:0.1:1 for Ca chips (Rare Metallic, 99.5%) and Ag (Kojundo Chemical, 99.9%), Pd (Kojundo Chemical, 99.9%), and black phosphorus (Kojundo Chemical, 99.9999%) powders were used for undoped and Pd-doped cases, respectively, which were put in alumina crucibles with Bi powder (Kojundo Chemical, 99.9%) and sealed in evacuated quartz tubes. The tubes were kept at 673 K for 12 h and 1273 K for 3 h and then slowly cooled to 673 K at a rate of 12 K h$^{-1}$, followed by centrifugation. Single crystals of CaAgAs were prepared by the method described in Ref. 19. As shown in the inset of Fig. 1(a), single crystals of CaAgP have a hexagonal columnar shape with widths of 0.1-0.2 mm, lengths up to 3 mm. The absence of Bi was confirmed by an energy dispersive X-ray spectrometer. The Pd content in the Pd-doped samples was estimated to be 0.24(14)% of the total molar ratio, which was much smaller than the nominal value, using an X-ray analysis microscope. This value yields $x = 0.007$ in CaAg$_{1-x}$Pd$_x$P. The electrical resistivity, magneto-resistance, Hall resistivity, and heat capacity were measured using a Physical Property Measurement System (Quantum Design).

## III. RESULTS AND DISCUSSION
### A. Transport properties

Figure 1(a) shows temperature dependence of electrical resistivity (ρ) of single crystals of undoped and Pd-doped CaAgP. The data of CaAgAs are also shown as reference. The undoped CaAgP and CaAgAs samples exhibited similar metallic temperature dependences. ρ of Pd-doped CaAgP was also metallic, but was different to the undoped samples in that ρ strongly decreased below 50 K. Similar behavior was also observed for Dirac semimetals such as $Cd_3As_2$ and $Na_3Bi$ [28-30]. As discussed later with the magnetoresistance data, this behavior is considered to be a manifestation of a weak antilocalization effect. Despite the strong decrease of ρ at low temperatures, the residual resistivity of the Pd-doped CaAgP was larger than that of the undoped samples, which is naturally considered as the disorder effect due to chemical doping.

Next, we discuss the magnetoresistances. First, the transverse magnetoresistance of the undoped CaAgP showed a concave-downward magnetic-field dependence at all temperatures measured, as shown in Fig. 1(b), which is different from the previous study using the polycrystalline samples [26]. Magnetoresistance at a magnetic field of 9 T was 15-20% of the zero-field ρ at any temperature, but the magnetic-field dependence in low magnetic field region became steeper with decreasing temperature. The magneto-resistance at 5 K exceeded 10% at 0.4 T, followed by a gradual increase at higher magnetic fields.

The magnetoresistance of the Pd-doped CaAgP is more extreme than that of the undoped sample. As shown in Fig. 1(d), transverse magnetoresistance at 5-50 K drastically increased below 0.1 T and showed an almost constant value above 1 T. This increase was quite strong, comparable to the increasing rate in high-quality single crystals of $Cd_3As_2$, but this magnetoresistance differs from $Cd_3As_2$ in that it saturates rapidly in a low magnetic field [28]. As seen in Fig. 1(d), the magnetic-field dependence became less prominent at higher temperatures. This anomalous magnetoresistance also appeared as a characteristic temperature dependence, as shown in Fig. 1(g). The strong decrease of the zero-field ρ below 50 K was almost completely suppressed by applying a magnetic field of 0.1 T.

The magnetoresistance of the Pd-doped CaAgP differed from that of the undoped sample also in terms of their magnetic-field orientation dependence. As shown in Fig. 1(e), the longitudinal magnetoresistance of the Pd-doped CaAgP steeply increases at low magnetic fields, similar to the transverse case, and reached an almost constant value of 0.04 mΩ cm above 1 T. This value was about half of the transverse magnetoresistance, as also seen in Fig. 1(f), suggesting that the Pd-doped sample intrinsically had sizable longitudinal magnetoresistance, in contrast to the undoped case shown in Fig. 1(c). In addition, the longitudinal magnetoresistance exhibited an oscillation above 1 T, which is likely due to quantum oscillations. This oscillation did not appear in the transverse magnetoresistance, suggesting that it corresponds to the orbit in the $k_x$–$k_y$ plane.

The strong decrease of the zero-field ρ below 50 K and the steep increase of magnetoresistance at low magnetic fields in the Pd-doped samples were most likely due to a reduction of back-scattering processes caused by quantum interference effects, i.e., weak antilocalization, and the destruction of it by applying a magnetic field [31-33]. The weak antilocalization is known to occur in metals with strong spin-orbit coupling [34], but this mechanism cannot explain the observed weak antilocalization behavior in the Pd-doped sample. This is because the magnitude of the decrease in electrical conductivity by applying a magnetic field, i.e., magneto-conductance, in the Pd-doped sample is more than two orders of magnitude larger than that expected by this mechanism. Another factor to consider is the impurity scattering in a nodal-line semimetal. It was pointed out that in the nodal-line semimetals, weak localization appears when the short-range impurity potential is dominant, while weak antilocalization appears when the long-range impurity potential is dominant [35]. The observed magnetoconductance is comparable to the calculated magnetoconductance when the latter case is realized in CaAgP, suggesting that the presence of a nodal line might play an important role in the emergence of weak antilocalization behavior in Pd-doped CaAgP. It is expected in future studies to quantitatively verify whether the observed weak antilocalization is due to the nodal line or not by precise measurements in a low magnetic field region at various temperatures.

Here we discuss the observed anomalous transport properties of CaAgP in terms of the properties of conducting carriers. As shown in Fig. 2(a), the undoped CaAgP showed positive and linear Hall resistivity ($ρ_H$), indicating that hole carriers were dominant. As shown in Fig. 2(c), hole carrier density was estimated to be $n_h = 4 \times 10^{19}$ cm$^{-3}$ at all temperatures measured, which was smaller than $n_h$ of the polycrystalline sample, indicating that the single crystal had fewer defects than the polycrystalline sample [26]. The mobility of hole carriers at 5-50 K was $μ_h = 1300$ cm$^2$ V$^{-1}$ s$^{-1}$, which is an order of magnitude larger than the polycrystalline case. By combining the $n_h$ value and the results of first-principles calculations in the same way as in Ref. 25, we found that $E_F$ for the undoped sample is located 0.14 eV below the Dirac point [Fig. 2(f)]. Since this energy is in a linearly dispersive region, a torus-shaped hole surface derived from the nodal ring was realized in the undoped CaAgP, as for the case of CaAgAs [6,19].

Unprecedentedly, the Pd-doped samples have electron carriers with extremely high mobility, in addition to hole carriers. As shown in Fig. 2(b), $ρ_H$ of the Pd-doped samples shows nonlinear behavior, indicating the coexistence of both electron and hole carriers in a sample. At 5 K, $ρ_H$ sharply decreases as the magnetic field increased from zero, shows a negative peak at 0.08 T, and linearly increases above this magnetic field. The negative $ρ_H$ clearly indicates the presence of electron carriers in the sample. The magnetic-field depend-ence becomes more gradual at higher temperatures. Figure 2(e) shows the hole conductivity ($\sigma_{xy}$) obtained from the $\rho_H$ and transverse magnetoresistance, and the results of fitting by $\sigma_{xy} = [n_h \mu_h^2/(1 + \mu_h^2 B^2) - n_e \mu_e^2/(1 + \mu_e^2 B^2)]eB$ to the data, where $n_h$ and $n_e$ are the carrier densities of the hole and electrons, $\mu_h$ and $\mu_e$ are the mobilities of the hole and electron carriers, respectively, and $B$ is the magnetic flux density [36]. The $\sigma_{xy}$ data at all measured temperatures were well reproduced by the above equation assuming the coexistence of hole and electron carriers in a sample, indicating the presence of hole carriers with $n_h$ and $\mu_h$ and electron carriers with $n_e$ and $\mu_e$. As shown in Fig. 2(c), $n_h$ is about $2 \times 10^{20}$ cm$^{-3}$ at all measured temperatures, which is larger than that of the undoped sample. By combining the $n_h$ value and the results of first-principles calculations for CaAgP, $E_F$ is located 0.4 eV below the Dirac point [Fig. 2(f)]. This value indicates that a torus-shaped hole surface related to the nodal ring is also formed in the Pd-doped sample, although it is thicker than that in the undoped case. As shown in Fig. 2(d), $\mu_h$ of the Pd-doped sample was an order of magnitude smaller than that of the undoped sample, which is natural to consider the larger $n_h$ and the disorder due to chemical doping.

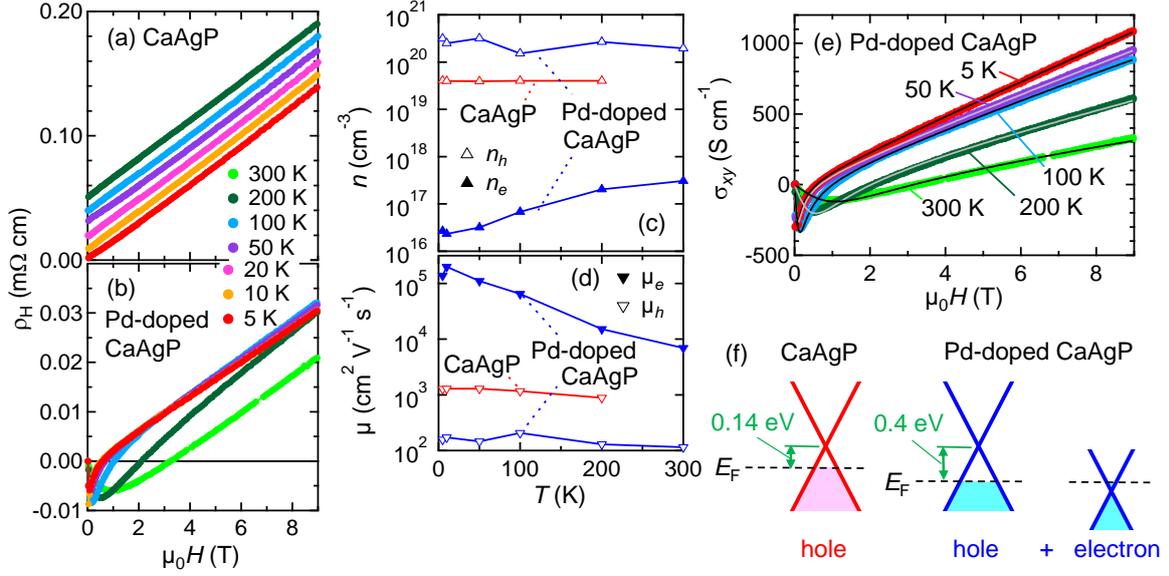

Fig. 2. (a,b) Hall resistivity of undoped and Pd-doped CaAgP single crystals measured at various temperatures. In (a), the data taken at 10-200 K have been shifted upward for clarity. (c,d) Carrier density and mobility of the undoped and Pd-doped CaAgP single crystals. The data of the undoped CaAgP were obtained by a linear fit of the Hall resistivity data, while those for Pd-doped one were obtained by fitting the Hall conductivity data shown in (e). (e) Hall conductivity of the Pd-doped CaAgP. The solid curves are the fitting results described in the main text. (f) Schematic pictures of the electronic states near $E_F$ in undoped and Pd-doped CaAgP.

In contrast, $n_e$ and $\mu_e$ at 10 K were $2.3 \times 10^{16}$ cm$^{-3}$ and $2.0 \times 10^5$ cm$^2$ V$^{-1}$ s$^{-1}$, respectively, indicating that there were a few electron carriers with extremely high mobility. This $\mu_e$ value is very large comparable to that of TaAs ($2$-$5 \times 10^5$ cm$^2$ V$^{-1}$ s$^{-1}$) [37,38], although smaller than the maximum values for Cd$_3$As$_2$ ($4 \times 10^6$ cm$^2$ V$^{-1}$ s$^{-1}$) and $\alpha$-(BEDT-TTF)$_2$I$_3$ ($8 \times 10^5$ cm$^2$ V$^{-1}$ s$^{-1}$) [28,39]. This means that the electron carriers in Pd-doped CaAgP reflect a Dirac-like dispersion, as similar to the carriers in the Dirac and Weyl semimetals. With increasing temperature, $n_e$ increases and $\mu_e$ decreases, as seen in Figs. 2(c) and 2(d), and $n_e$ and $\mu_e$ at 300 K are an order of magnitude larger and smaller than those at 5-10 K, respectively.

The above results indicate that Pd doping to CaAgP leads to the emergence of electron carriers exhibiting the characteristics of Dirac electrons as well as hole doping by Pd substitution to the Ag site [Fig. 2(f)]. Assuming a rigid-band picture, substitution of Pd atoms to the Ag site is a hole doping, because a Pd atom has one less electron than a Ag atom. Given that one hole carrier is generated by the substitution of one Pd atom, the Pd proportion of 0.24% corresponds to a hole carrier density of $1.3 \times 10^{20}$ cm$^{-3}$. This value is comparable to $1.6 \times 10^{20}$ cm$^{-3}$, that is the difference between the $n_h$ values of the Pd-doped and undoped samples, suggesting that the doped Pd atoms acted as an acceptor. In contrast, the emergence of electron carriers cannot be explained by such a rigid-band picture for a bulk band. A possible scenario explaining the emergence of electron carriers is the effect of a topological surface band associated with the nodal ring, such as the situation where the surface band crosses the $E_F$. Other possibilities are the effects of disorder and microscopic phase separation caused by the Pd doping, but it is not easy to realize high mobility by these effects. It is still unclear how such high-mobility electron carriers appeared. It would be interesting to clarify the formation mechanism of the electron carriers in future experiments using microscopic probes.

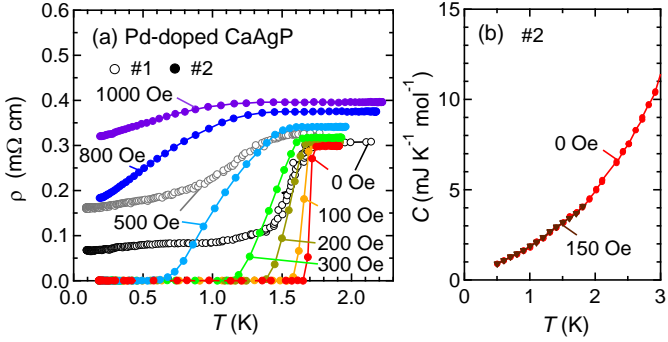

Fig. 3. Temperature dependences of electrical resistivity (a) and heat capacity (b) of the single crystals of Pd-doped CaAgP measured at various magnetic fields.

### B. Superconductivity in the Pd-doped samples

Finally, we report on a superconducting transition observed in the Pd-doped samples. Figure 3(a) shows ρ of Pd-doped samples of #1 and #2. Sample #1 was synthesized under the same condition as those shown in Figs. 1 and 2, while #2 was synthesized using Pd at twice the molar ratio employed for the sample #1. With decreasing temperature, ρ of the sample #1 strongly decreases below 1.70 K. ρ of sample #2 also shows a sharp drop below 1.75 K and became zero at 1.65 K. As shown in Fig. 3(a), the temperatures at which sharp drops appear decreased by applying a magnetic field, suggesting that the resistivity drops in both samples are caused by a superconducting transition. As shown in Fig. 3(b), however, heat capacity (C) of the sample #2 showed no (or a negligibly small) anomaly corresponding to the superconducting transition, indicating that the observed superconductivity was not simple bulk superconductivity. An interesting scenario explaining the behaviors of ρ and C is surface superconductivity due to the topological surface band associated with the nodal ring. Unlike the point-node case, materials with a nodal ring can have a flat surface band, which gives rise to large density of states. In the Pd-doped samples, the surface states related to the presence of high-mobility electron carriers are possibly formed near $E_F$, which may be responsible for the observed superconductivity. At present, however, there is no direct evidence of surface superconductivity. Since other scenarios, such as bulk superconductivity due to a phase with a very small Sommerfeld coefficient or an impurity phase, are also assumed, it is important to confirm which scenario is responsible for the superconductivity in Pd-doped CaAgP in the future experiments using surface-sensitive techniques.

### IV. CONCLUSION

In conclusion, we studied the transport properties of single crystals of a candidate nodal-line semimetal CaAgP. The transport properties of undoped CaAgP are understood within the framework of a hole-doped nodal-line semimetal. In contrast, the Pd-doped samples have not only hole carriers due to the Pd doping, but also electron carriers with extremely high mobility, which are expected to exist in proximity of the Dirac point. The pronounced weak antilocalization effect in the transport properties, such as the drastic increase of magnetoresistance at low magnetic fields and the strong decrease of electrical resistivity at low temperatures, are most likely due to the high-mobility electron carriers. Electrical resistivity of the Pd-doped samples further showed a superconducting transition with onset temperatures of 1.7-1.8 K. These results suggest that CaAgP realizes intriguing electronic properties unique to nodal lines.

### ACKNOWLEDGMENTS

The authors are grateful to H. T. Hirose for helpful discussions. This work was partly carried out at the Materials Design and Characterization Laboratory under the Visiting Research Program of the Institute for Solid State Physics, University of Tokyo and partly supported by the Collaborative Research Project of Materials and Structures Laboratory, Tokyo Institute of Technology, JSPS KAKENHI (Grant Nos. 19K21846, 19H05823, and 20H02603), and CREST, JST (Grant No. JPMJCR1421).